# Emergence of magnetism and controlling factors of superconductivity in Li/Na-ammonia co-intercalated FeSe$_{1-z}$Te$_z$

*Hechang Lei,*[1,†,§] *Jiangang Guo,*[1,†] *Fumitaka Hayashi,*[2] *and Hideo Hosono,*[1,2,3*]

[1] Materials Research Center for Element Strategy, Tokyo Institute of Technology, Yokohama 226-8503, Japan

[2] Frontier Research Center, Tokyo Institute of Technology, Yokohama 226-8503, Japan

[3] Materials and Structures Laboratory, Tokyo Institute of Technology, Yokohama 226-8503, Japan

## Abstract

The discovery of superconductivity in alkali-ammonia co-intercalated FeSe has generated intensive interest because of their highest $T_c$ (~ 45 K) among iron-chalcogenide superconductors with bulk form. Here, we report the phase diagrams of two series of Li/Na-ammonia co-intercalated FeSe$_{1-z}$Te$_z$. When superconductivity is suppressed by Te doping, the magnetic ordering states are emergent. Moreover, a novel phase Li$_x$(NH$_3$)$_y$Fe$_{2-\delta}$Te$_2$ with possible antiferromagnetism is discovered. This strongly indicates the intimate relation between superconductivity and magnetism in these materials, like in other iron-based superconductors. On the other hand, comparative structure analysis with FeSe$_{1-z}$Te$_z$ suggests that although there is remarkable similarity in phase diagrams for both iron-chalcogenide and iron-pnictide superconductors, the different types of anions with different charges lead to the dissimilar controlling factors in superconductivity for both classes of materials. This opens up an opportunity to tune the superconductivity in iron-chalcogenide superconductors in more ways than just one.

[†]These authors contributed equally to this work.

[§]Present address: Department of Physics, Renmin University of China, Beijing 100872, China

[*]To whom correspondence may be addressed. E-mail: hosono@msl.titech.ac.jp

**Introduction**

The superconductivity of FePn-based superconductors (SCs) is usually in the vicinity of magnetic ordering states [1-3], leading to a possible unconventional superconducting mechanism and intricate interplay (competition and/or coexistence) between superconductivity and magnetism [4, 5]. Similarly, the superconducting states in FeCh-based (Ch = S, Se, and Te) SCs are also near the magnetic ordering states [6-10], such as the emerging spin glass (SG) state in $K_xFe_{2-y}(Se, S)_2$ with S doping [8], the antiferromagnetism (AFM) with ($\pi$, 0) in-plane magnetic wave vector in FeTe [9] and the recently discovered antiferromagnetic and semiconducting/insulating state in $\beta$-$Fe_{1-x}$Se with particular Fe-vacancy orders [10].

On the other hand, after summarizing a large amount of experimental data, the superconducting transition temperature $T_c$ of FePn-based SCs is closely related to two structural parameters, the bond angle ($\alpha$ and $\beta$) of Pn-Fe-Pn (Pn = P and As) and the anion height from the Fe layer ($h$) [11, 12]. In general, $T_c$ increases when $\alpha$ and $\beta$ approach the angle of a regular tetrahedron (109.47º), or $h$ is close to the optimal value (~ 1.38 Å). These correlations can partially explain the trends in $T_c$ for some FeCh-based (Ch = S, Se, and Te) SCs, such as the enhancement of $T_c$ in FeSe with pressure that could be related to the decrease of $h$ to the optimal value [13, 14], and the decrease of $T_c$ in $K_xFe_{2-y}(Se, S)_2$ that is ascribed to the increase of bond angles of Ch-Fe-Ch [8]. However, these are many exceptions. For example, even though the anion height of intercalated FeSe is far from the optimal value, its $T_c$ (~ 30 K) is still much larger than that of FeSe [15]. The $T_c$ of FeSe with better $\alpha/\beta$ and $h$ is lower than that of $FeTe_{0.53}Se_{0.47}$ [11, 12]. These violations should be closely related to the remarkably different anions with different charges, $Ch^{2-}$ vs. $Pn^{3-}$. Thus, the question of whether the $T_c$ is influenced by the same factors in FePn- and FeCh-based SCs is still under debate.

Recently, superconductivity with $T_c$ up to 45 K has been reported in $A_x(NH_3/NH_2^-)_yFe_2Se_2$ (A = alkali, alkali-earth and rare-earth metals) synthesized by the low-temperature ammonothermal method [16-20]. Until now, it is still unknown whether this newly discovered superconducting family has a similar magnetic/superconducting phase diagram as in other iron-based SCs, which is vitally important to understanding the relationship between superconductivity and magnetism in these materials. On the other hand, this system is a good platform to study the relationship between structure and superconductivity in FeCh-based SCs because the FeSe layer is almost intact (few Fe vacancy) even after intercalation. This avoids the complexities of structure and property found in high-temperature-synthesized $A_x$Fe$_{2-\delta}$Ch$_2$ (A = K, Rb, Cs, and Tl) due to the formation of a large amount of Fe vacancies [21-23]. Here, we systematically studied the phase diagrams of two series of Li$_x$(NH$_3$)$_y$Fe$_{2-\delta}$(Se$_{1-z}$Te$_z$)$_2$ and Na$_x$(NH$_3$)$_y$Fe$_{2-\delta}$(Se$_{1-z}$Te$_z$)$_2$ ($0 \leq z \leq 1$) (abbreviated as Li/NaNH$_3$-122). The detailed magnetic characterization indicates that the superconducting states in these compounds are close to the magnetic ordering states. On the other hand, the comparative structure analysis with FeSe$_{1-z}$Te$_z$ indicates that the controlling factors in $T_c$ in FeCh-based SCs are different from those in FePn-based SCs due to the different chemical features of anions with different charges.

**Experimental**

The highly-pure powder Fe(Se,Te) precursors were synthesized using solid-state reactions. The iron granules (Alfa, 99.98%), selenium grains (Kojundo, 99.99%) and Te grains (Kojundo, 99.999%) were placed into alumina crucibles and sealed in silica ampoules. The samples were heated to 1323 K for 30 h and kept at 1323 K for 24 h, then annealed at 673 K for 50 h, and finally furnace-cooled to room temperature. The obtained Fe(Se,Te) parent compounds were ground into powder and loaded into a Taiatsu Glass TVS-N$_2$ high-pressure vessel (30 ml) with alkali-metal pieces (Li/Na : Fe(Se,Te) = 1 : 2 molar ratio). A magnetic stirrer was also loaded and the vessel was closed. All of above processes were carried out in an argon-filled glove box with an O$_2$ and H$_2$O content below 1 ppm. The vessel was taken out from the glove box and connected to a vacuum/NH$_3$ gas line equipped with a molecular pump and mass-flow controller. Before introducing NH$_3$, the vessel was evacuated using a molecular pump ($\sim 5\times10^{-2}$ Pa) and placed in a bath of ethanol cooled by liquid nitrogen ($\sim$ 223 K).

Then, the ammonia cylinder and regulator were then opened and 5 - 10 g of $NH_3$ was condensed into the vessel. After that, the reaction vessel was closed and stirred for ~ 3 - 10 h at 223 - 243 K. When the intercalated process was finished, the vessel was opened and the solutions were evaporated at an ambient pressure. For the Li and $NH_3$ co-intercalated samples, the vessel was further evacuated to ~ $10^{-2}$ Pa using a molecular pump for ~ 0.5 h in order to remove the $NH_3$-rich phase in the samples [19].

The powder X-ray diffraction (PXRD) patterns of products were measured by a Bruker diffractometer model D8 ADVANCE with Mo-$K_\alpha$ radiations ($\lambda$ = 0.7107 Å) at room temperature. The samples were loaded into thin-walled capillary tubes (diameter = 0.5 mm) and then rotated during the measurement in order to eliminate the preferred orientation of samples. The Rietveld refinement of patterns was performed using code TOPAS4 (Bruker AXS). Magnetic susceptibility was measured by a vibrating sample magnetometer (SVSM, Quantum Design). The chemical compositions of the samples were determined by energy-dispersive X-ray analysis (EDX). The compositions were determined as the average values of 20 points. The contents of nitrogen in the samples were determined using an ion chromatography (IC) technique [24]. Typically, 10 mg of the sample was dissolved in 5 mol/L HF aqueous solution, and was diluted by adding water. The resultant solution was analyzed by IC with a Shimadzu CDD-10A conductivity detector.

**Results and Discussion**

Figure 1(a) and (b) show the $c$- and $a$-axial lattice parameters of Li/NaNH$_3$-122 as determined from the powder X-ray diffraction (PXRD) (Fig. S1, Supplemental Material [25]) using the structure of $Li_{0.6}(ND_{2.8})Fe_2Se_2$ (Fig. 1(c)) as an initial model [17]. The fitted lattice parameters of $FeSe_{1-z}Te_z$ from PXRD patterns (Fig. S2, Supplemental Material [25]) are also plotted for comparison. The $a$- and $c$-axial lattice parameters for both Li/NaNH$_3$-122 and $FeSe_{1-z}Te_z$ increase almost linearly with Te doping, confirming the substitution of Se by larger Te. It should be noted that the coexistence of two phases in $FeSe_{1-z}Te_z$ with $0.1 \leq z \leq 0.5$ (Fig. S2, Supplemental

Material [25]) leads to similar behaviors in the intercalated compounds (Fig. 1(a) and (b)), which are different from samples synthesized using a high-temperature method where all of the starting materials are fused together [23]. In the two-phase region, one phase (Phase 1) has smaller $a$- and $c$-axial lattice parameters than the other one (Phase 2), possibly because of the lower Te content in Phase 1 than in Phase 2. Although the $c$-axial lattice parameters of Li/NaNH$_3$-122 are remarkably larger than those of FeSe$_{1-z}$Te$_z$ due to the co-intercalation with NH$_3$ and NH$_2^-$ (Fig. 1(a)), the relative increase [$(d_{(00l), z=0} - d_{(00l), z=0.9})/d_{(00l), z=0}$] of Li/NaNH$_3$-122 (~ 11.6 % and 7.8 %, respectively) is smaller than that in FeSe$_{1-z}$Te$_z$ (~ 13.6 %). It can be ascribed to the less change of thick Li-/Na-NH$_3$(NH$_2^-$) layers with Te doping that weakens the influence of the change of Fe(Se, Te) layer on the $c$-axial lattice parameters. The fact that the $a$-axial lattice parameters of Li/NaNH$_3$-122 are slightly larger than those of FeSe$_{1-z}$Te$_z$ at certain $z$ values indicates that the Fe(Se, Te) layers are stretched after intercalation (Fig. 1(b)). The relative change of the $a$-axial lattice parameters is ~ 1.8 % and 1.9 % for LiNH$_3$-122 and NaNH$_3$-122, respectively, close to the value of ~ 1.4 % in FeSe$_{1-z}$Te$_z$ but much smaller than those values for the $c$ axis. It suggests that the interlayer interaction between the Fe(Se, Te) layers is much weaker than intralayer one. Interestingly, for FeTe, Li-NH$_3$ can be co-intercalated and the PXRD pattern fits well using the isostructure of Li$_{0.6}$(ND$_{2.8}$)Fe$_2$Se$_2$ (Fig. 1(d)). In contrast, Na-NH$_3$ cannot be co-intercalated into FeTe completely (Fig. S3, Supplemental Material [25]), unlike the situation in the counterpart FeSe. This could be due to the smaller ionic size of Li$^+$ than Na$^+$ and the different chemical features of FeTe and FeSe.

The chemical composition analyses of Li/NaNH$_3$-122 (Table S1, Supplemental Material [25]) show that the actual molar ratios of Se to Te are comparable to those in FeSe$_{1-z}$Te$_z$ (Table S2, Supplemental Material [25]). On the other hand, there is no systematic change of Na content with Te doping, which fluctuates between ~ 0.45 and 0.8. These values are close to the previous results for (Li/Na/K)$_x$(NH$_3$)$_y$Fe$_{2-\delta}$Se$_2$ [17-20]. Ion chromatography (IC) analysis indicates that the NH$_3$ ($y$) content for the whole series ranges between ~ 0.3 and 0.8, close to the results reported in the literature [17, 18, 20], assuming that all nitrogen species originate from NH$_3$ molecules. Most strikingly, the Fe vacancies (~ 1 - 10 %, see Table S1 in Supplemental Material [25]) are much lower than those of high-temperature-synthesized $A_x$Fe$_{2-\delta}$Ch$_2$ (~ 10 - 25 %) [21, 22], which has significant effects on the magnetism of samples at the normal state discussed below.

As shown in Figure 2(a) and (b), the onset of $T_c$ is 41.4 K and 45.7 K for (Li/Na)$_x$(NH$_3$)$_y$Fe$_{2-\delta}$Se$_2$, respectively, consistent with the reported results [16, 17, 20]. With increasing Te content, the $T_c$ shift to the lower temperature and there are two transitions for $0.2 \leq z \leq 0.4$, due to the coexistence of two phases. When doping Te further ($z \geq 0.5$), the two transitions converge again, with even lower $T_c$s. Finally, the superconducting transition cannot be observed above 2 K for both LiNH$_3$-122 and NaNH$_3$-122 with $z \geq 0.8$. Except for the sample with $z = 0.7$, all of the superconducting samples exhibit large superconducting volume fractions (SVFs), implying the superconductivity should be bulk. In order to confirm that the absence of superconductivity in the samples with $z \geq 0.8$ is not due to the low SVFs in FeSe$_{1-z}$Te$_z$ ($z \geq 0.8$) [26], a Na-NH$_3$ co-intercalated sample using O$_2$-annealed FeSe$_{0.2}$Te$_{0.8}$ was prepared. The magnetic susceptibility measurements indicate that even if the O$_2$-annealed FeSe$_{0.2}$Te$_{0.8}$ shows bulk superconductivity, the superconductivity is still absent for the intercalated sample (Fig. S4, Supplemental Material [25]). Thus, the absence of superconductivity for the samples with $z \geq 0.8$ should be intrinsic, irrespective of the status of parent compounds.

At high temperature, the magnetic susceptibilities of all samples for both series decrease with increasing temperature at $H$ = 1 kOe. Typical curves are shown in Fig. 3(a) and Fig. S5(a) in Supplemental Material [25]. They are remarkably different from those of high-temperature-synthesized $A_x$Fe$_{2-\delta}$Ch$_2$ in which positive or weak temperature dependence of $\chi(T)$, are often observed up to room temperature [8, 27]. This distinctive behavior should be attributed to the good integrity of the Fe(Se, Te) layers with fewer Fe vacancies in Li/NaNH$_3$-122 and the absence of AFM above room temperature as in $A_x$Fe$_{2-\delta}$Ch$_2$ because of the Fe vacancies ordering. At the high-Te-doping region, there is a magnetic transition appearing below $T_p$ (Fig. 3(a) and Fig. S5(a) in Supplemental Material [25]). The bifurcation of ZFC and field-cooling (FC) curves at $T_p$ suggests the presence of a spin-glass (SG) transition. But the detailed measurements indicate that they are different for samples with different Te contents. For (Li/Na)$_x$(NH$_3$)$_y$Fe$_{2-\delta}$(Se$_{0.1}$Te$_{0.9}$)$_2$, the transitions exhibit strong field dependence even at high field regions (Fig. 3(b) and Fig. S5(b) in Supplemental Material [25]). But for Li$_x$(NH$_3$)$_y$Fe$_{2-\delta}$Te$_2$, although the transition shows field dependence at very low fields, which could be due to the effect of possible ferromagnetic impurities in the sample such as deintercalated free Fe-ions/atoms, the $T_p$ is insensitive to the field when $H \geq 1$ kOe (Fig. 3(c)). On the other hand, for Li$_x$(NH$_3$)$_y$Fe$_{2-\delta}$(Se$_{0.1}$Te$_{0.9}$)$_2$, the $T_p$ of the real part of the ac susceptibility $\chi'(T)$ shifts to higher temperature with higher frequency, $f$ (Fig. 3(d)) and the fitted $K$ [= $\Delta T_p/(T_p \Delta \log f)$] is 0.0148(9) (Fig. 3(f)). It is consistent with the values (0.0045 ≤ $K$ ≤ 0.08) found in the canonical SG system, but much smaller than that in a superparamagnet [28]. Similar behavior is also observed in Na$_x$(NH$_3$)$_y$Fe$_{2-\delta}$(Se$_{0.1}$Te$_{0.9}$)$_2$ (Fig. S5(c) and (c), Supplemental Material [25]). In contrast, there is almost no shift of $T_p$ on $\chi'(T)$ curves with frequency in Li$_x$(NH$_3$)$_y$Fe$_{2-\delta}$Te$_2$ sample (Fig. 3(e) and (f)). These results suggest that the magnetic transitions for Li/NaNH$_3$-122 (0.6 ≤ $z$ ≤ 0.9) should be SG transitions but that Li$_x$(NH$_3$)$_y$Fe$_{2-\delta}$Te$_2$ might have an antiferromagnetic transition. It should be noted that the bifurcation of ZFC and FC curves below antiferromagnetic transition in Li$_x$(NH$_3$)$_y$Fe$_{2-\delta}$Te$_2$ and the difference of $T_p$s between dc and ac susceptibilities in Li/Na$_x$(NH$_3$)$_y$Fe$_{2-\delta}$(Se$_{0.1}$Te$_{0.9}$)$_2$ could also be ascribed to the existence of ferromagnetic impurities. For Li$_x$(NH$_3$)$_y$Fe$_{2-\delta}$Te$_2$, $T_p$ is ~ 30 K at $H = 1$ kOe (Fig. 3(c)), lower than the antiferromagnetic transition of FeTe with $T_N$ ~ 60 - 75 K [9]. However the SG states in Li/NaNH$_3$-122 (0.6 ≤ $z$ ≤ 0.9) are more robust than that in FeSe$_{1-z}$Te$_z$, where the antiferromagnetic state is suppressed significantly when ~ 10 % Se is incorporated into the Te lattice [9].

The phase diagrams of $T_c(z)$ and $T_p(z)$ for Li/NaNH$_3$-122 are summarized in Fig. 4. There are two $T_c$s ($T_{c1}$ and $T_{c2}$) in a two-phase region, similar to FeSe$_{1-z}$Te$_z$ (0.1 ≤ $z$ ≤ 0.5) (Fig. S6(b) in Supplemental Material [25]), of which the $T_c$ values are determined from $4\pi\chi(T)$ curves (Fig. S6(b) in Supplemental Material [25]). The $T_c$s decrease monotonically with Te doping and show similar values for both series at a certain $z$ value, indicating the $T_c$ is not sensitive to the kind of intercalated alkali metals. It could be ascribed to the existence of $NH_3$ molecules which weakens the direct interaction of alkali metal ions on Fe(Se, Te) layers and leads to the insensitivity of superconductivity to the ionic type. The suppression of $T_c$s is similar to that of $K_xFe_{2-y}(Se_{1-z}S_z)_2$ [12], but much milder than that of $Rb_{0.8}Fe_{2-y}(Se_{1-z}Te_z)_2$ where doping 15% Te into Se site has already quenched the superconductivity [23]. For $z \geq 0.6$, there are magnetic transitions that change from SG transition to an antiferromagnetic one when all of Se is replaced by Te. All samples exhibit paramagnetism at high temperatures. It should be noted that the spin glass or antiferromagnetic transition temperatures $T_p$s shown in the phase diagram are determined from $\chi(T)$ curves at $H = 1$ kOe, which is lower than those $T_p$s on $\chi'(T)$ curves with $H_{ac} = 1$ Oe, because the existence of trace of ferromagnetic impurities might influence the $T_p$s significantly at low field. Moreover, whether the coexistence of SG state and superconductivity for $0.6 \leq z < 0.8$ is intrinsic or extrinsic due to the possible inhomogeneity is still an open question at present because of the difficulties characterizing of homogeneity of samples in that region. But in any case the trend towards suppression of superconductivity and emergence of magnetic order states with Te doping are undoubted. The fact that these phase diagrams of $Li/NaNH_3$-122 are similar to those of other iron-based SCs, undoubtedly originates from the same key element in these materials, Fe.

On the other hand, the structural analysis indicates that the changes of $T_c$ are intimately related to the structural evolution with Te doping. As shown in Fig. 5(a) and (b), in general, the bond angles of Ch-Fe-Ch ($\alpha$ and $\beta$) deviate from the 109.47º of an ideal tetrahedron, and the anion height $h$ that is always larger than 1.38 Å also increases with Te doping for $Li/NaNH_3$-122 and $FeSe_{1-z}Te_z$. Both trends agree with the empirical relations between structure and $T_c$ mentioned above [11, 12]; therefore, the suppression of $T_c$ in $Li/NaNH_3$-122 could be ascribed to increasing distortion of the FeCh tetrahedron and moving of anions away from the Fe plane. These parameters are related to each other by the equations: $h = d_{Fe-Pn}\cos(\alpha/2)$, and $\sin(\beta/2)/\sin(\alpha/2) = [1/2+(2w-1/2)^2(c/a)^2]^{1/2}$ for $Li/NaNH_3$-122 ($= [1/2+4w^2(c/a)^2]^{1/2}$ for $FeSe_{1-z}Te_z$), where $d_{Fe-Pn}$ is the Fe-Pn bond length, $w$ is the $c$-axial fractional coordination, and $c$ and $a$ are lattice parameters for $c$ and $a$ axes; therefore, it is difficult to distinguish which parameter is dominant in $T_c$. On the other hand, these empirical relations cannot explain the remarkably low $T_c$ in FeSe$_{1-z}$Te$_z$ even with similar $\alpha/\beta$ and $h$ to those of Li/NaNH$_3$-122 and the higher $T_c$ in FeSe$_{1-z}$Te$_z$ with Te doping along with the larger distortion of the Fe-Ch tetrahedron and $h$.

Besides $\alpha/\beta$ and $h$, the interlayer distance $d$ between FeCh layers is also considered a key parameter influencing $T_c$ [29, 30]. The $T_c$ increases with increasing $d$ and then becomes saturated at ~ 45 K above 9 Å [29, 30]. This correlation between $d$ and $T_c$ has also been reported in the high-temperature synthesized K$_x$Fe$_{2+\delta}$Se$_2$ ($T_c$ = 44 K) with much larger $d$ (= 9.05 Å) and excess Fe than extensively studied K$_x$Fe$_{2-\delta}$Se$_2$ ($T_c$ ~ 30 K) with $d$ ~ 7 Å and ordered vacancy [31]. A similar trend exists in $\beta$-HfNCl. The increase of $T_c$ with $d$ is attributed to the enhancement of Fermi surface (FS) nesting, originating from the reduced dispersion along the $c$ axis, i.e., the less warps of FSs [32, 33]. The saturation with slowly decreasing $T_c$ of electron-doped $\beta$-HfNCl for $d$ > 15 Å might be coupled with a Coulomb interaction between the layers [32]. Thus, the weakening effect of $d$ on the warp of FSs will diminish when its value is beyond certain threshold. As shown in Fig. 5(c), the $d$ values increase monotonically with increasing Te content for all three series, and the $d$ values of Li/NaNH$_3$-122 are remarkably larger than those of FeSe$_{1-z}$Te$_z$. Because the $d$ values of FeSe$_{1-z}$Te$_z$, are small, the parameter $d$ may play a more important role in influencing $T_c$ than $\alpha/\beta$ and $h$. But for Li/NaNH$_3$-122, the $d$ values are close to 9 Å and its effect on $T_c$ might become weak. Other factors, such as $\alpha/\beta$ and $h$, etc., will dominate $T_c$.

Although the degree of charge-doping may be not the dominant factor on $T_c$ for Li/NaNH$_3$-122, it has to be considered as the third parameter influencing $T_c$ of FeCh-based SCs since K$_{0.6}$(NH$_3$)$_y$Fe$_2$Se$_2$ and K$_{0.3}$(NH$_3$)$_y$Fe$_2$Se$_2$ show different $T_c$s at 30 K and 44 K [18]. But the degree of charge-doping could not determine the $T_c$ solely because Li-intercalation into FeSe by the electrochemical method does not enhance $T_c$ significantly [34], in contrast to the Li/NH$_3$ co-intercalation.

Therefore, although the $T_c$ may be controlled by $\alpha/\beta$ and $h$ for Li/NaNH$_3$-122, the factors influencing $T_c$ are different for FePn- and FeCh-based SCs. $T_c$ is sensitive to more factors in FeCh-based SCs than in FePn-based ones and these factors are closely related to the electronic structure. They include (1) the interlayer distance that could control the dispersion of FSs along the $k_z$ direction; (2) the bond angles of Ch-Fe-Ch and the anion height that could determine the shape of FSs in the $k_x$ - $k_y$ plane; and (3) the degree of charge-doping that is related to the position of Fermi energy level. This conclusion could explain the trend of $T_c$s and resolve the apparent contradictions observed in FeCh-based SCs. The highest $T_c$ should result not from optimizing one of these parameters but from an optimal combination, i.e., a larger $d$ with more regular FeCh tetrahedron and better $h$ closer to the optimal value (~ 1.44 Å, not necessarily the same as the optimal one for FePn-based SCs) as well as a close-to-optimal degree of charge-doping. It is different from FePn-based SCs where superconductivity is usually determined by a single parameter ($\alpha/\beta$ or $h$). This difference should originate from the distinctively chemical features of anions with different charges in these two classes of materials. For FePn-based SCs, the FePn layers must be negatively charged and thus the blocking layers with positive valence are required. This necessity may limit the effects of the charge-doping level and interlayer distance on $T_c$. In contrast, for FeCh-based SCs, the neutrality of the FeCh layers relaxes these restrictions and these factors could play important roles in determining the $T_c$. This difference has already led to the realization of FeSe monolayer with significantly high $T_c$ [35], in contrast to the unattainably superconducting FeAs monolayer.

All of these factors have crucial influences not only on the superconductivity but also magnetism of iron-based materials. The lower $T_N$ in Li$_x$(NH$_3$)$_y$Fe$_{2-\delta}$Te$_2$ than and the robust SG states in Li/NaNH$_3$-122 ($0.6 \leq z \leq 0.9$) than those in FeSe$_{1-z}$Te$_z$ should be closely related to the changes of magnetic interaction because Li/Na-NH$_3$ co-intercalation induces more carriers and/or changes the structural parameters, such as stretching the Fe(Se, Te) plane and increasing the interlayer distance. More importantly, because of the distinctly structural differences of FeCh-based and FePn-based SCs, the factors controlling the magnetism of these two families should also be different. In a word, the structural factors are the most important and apparent factors for physical properties and the evolution of underlying various microscopic interactions that intimately connect to these structural factors controls the evolution of physical properties and leads to the variety of phase diagrams where different ordering states such as magnetism and superconductivity compete and/or coexist each other.

## Conclusions

In summary, we studied the phase diagrams of two series of Li/NaNH$_3$-122 with Te doping systematically. We found the magnetic ordering states are in the vicinity of the superconducting state in Li/NaNH$_3$-122. This universality strongly indicates the intimate relation between superconductivity and magnetism, suggesting the possible existence of an unconventional superconducting mechanism in Li/NaNH$_3$-122. It definitely originates from one key ingredient, Fe. In contrast to this universality, the factors influencing $T_c$ are different for both FePn- and FeCh-based SCs. $T_c$ is sensitive to more factors in FeCh-based SCs than in FePn-based ones, possibly because of the distinctly different chemical features of another key ingredient, anions (Ch$^{2-}$ vs. Pn$^{3-}$) It suggests that there are several ways to control superconductivity in FeCh-based SCs. These findings will not only shed light on exploring new iron-based SCs, especially FeCh-based SCs, with the higher $T_c$, but also help us to understand the relationship among structure, electronic structure, superconductivity and magnetism.

## Acknowledgements

This work was supported by the Funding Program for World-Leading Innovative R&D on Science and Technology (FIRST) and MEXT Element Strategy Initiative to form a core research center, Japan.## References

**Figure captions**

Figure 1. The fitted (a) $c$-axial and (b) $a$-axial lattice parameters for $(Li/Na)_x(NH_3)_yFe_{2-\delta}(Se_{1-z}Te_z)_2$ and $FeSe_{1-z}Te_z$. All the error bars are statistical errors from the TOPAS refinement. (b) The crystal structure of $(Li/Na)_xN_yFe_{2-\delta}(Se_{1-z}Te_z)_2$. Because the position of H atoms cannot be determined due to the smaller scattering factor of hydrogen, the H atoms are not plotted in the crystal structure. (b) The PXRD pattern of $Li_x(NH_3)_yFe_{2-\delta}Te_2$ .

Figure 2. Temperature dependences of magnetic susceptibility $4\pi\chi(T)$ for (a) $Li_x(NH_3)_yFe_{2-\delta}(Se_{1-z}Te_z)_2$ and (b) $Na_x(NH_3)_yFe_{2-\delta}(Se_{1-z}Te_z)_2$ at low-temperature region with $H = 10$ Oe. For clarity, only the zero-field-cooling (ZFC) curves are shown.

Figure 3. (a) $\chi(T)$ curves of $Li_x(NH_3)_yFe_{2-\delta}(Se_{1-z}Te_z)_2$ ($z = 0.4$, 0.7 and 0.8) at $H = 1$ kOe with ZFC and field-cooling (FC) modes. (b) and (c) Temperature dependence of magnetic moment $M(T)$ at various fields for $Li_x(NH_3)_yFe_{2-\delta}(Se_{1-z}Te_z)_2$ with $z = 0.9$ and $z = 1.0$, respectively. For $Li_x(NH_3)_yFe_{2-\delta}Te_2$, the kinks at $T \sim 68$ K with $H \geq 10$ kOe are due to the minor second phase of FeTe. (d) and (e) The real part of ac susceptibility $\chi'(T)$ at $H_{ac} = 1$ Oe with various frequencies $f$ for $Li_x(NH_3)_yFe_{2-\delta}(Se_{1-z}Te_z)_2$ with $z = 0.9$ and $z = 0.1$, respectively. The arrows show the $T_p$s on $\chi'(T)$ curves. (f) Frequency dependence of peak positions $T_p$s for $Li_x(NH_3)_yFe_{2-\delta}(Se_{1-z}Te_z)_2$ with $z = 0.9$ and $z = 1.0$. The red solid line is the result of a linear fit.

Figure 4. Superconducting and magnetic phase diagrams of $(Li/Na)_x(NH_3)_yFe_{2-\delta}(Se_{1-z}Te_z)_2$. The $T_p$s are determined from $\chi(T)$ curves at $H = 1$ kOe. The red solid and empty squares represent two superconducting transition

temperatures $T_{c1}$ and $T_{c2}$ of $Li_x(NH_3)_yFe_{2-\delta}(Se_{1-z}Te_z)_2$. The blue solid and empty circles represent $T_{c1}$ and $T_{c2}$ of $Na_x(NH_3)_yFe_{2-\delta}(Se_{1-z}Te_z)_2$. The green solid squares/violet solid circles represent the spin glass transition temperatures $T_p$s of $(Li/Na)_x(NH_3)_yFe_{2-\delta}(Se_{1-z}Te_z)_2$. The green empty square represents the antiferromagnetic transition temperature $T_p$ of $Li_x(NH_3)_yFe_{2-\delta}Te_2$. PM, paramagnetism; SC-Phase 1, superconductivity of Phase 1; SC-Phase 2, superconductivity of Phase 2; SG, spin glass; AFM, antiferromagnetism.

Figure 5. (a) Ch-Fe-Ch bond angles $\alpha$ and $\beta$, (b) anion height $h$, and (c) interlayer distance $d$ between Fe-Ch layers of $(Li/Na)_x(NH_3)_yFe_{2-\delta}(Se_{1-z}Te_z)_2$ and $FeSe_{1-z}Te_z$ as a function of $z$. Insets of (a), (b) and (c) show the definitions of these parameters in the structure. The dotted pink line in (a) shows the bond angle of a regular FeCh tetrahedron.

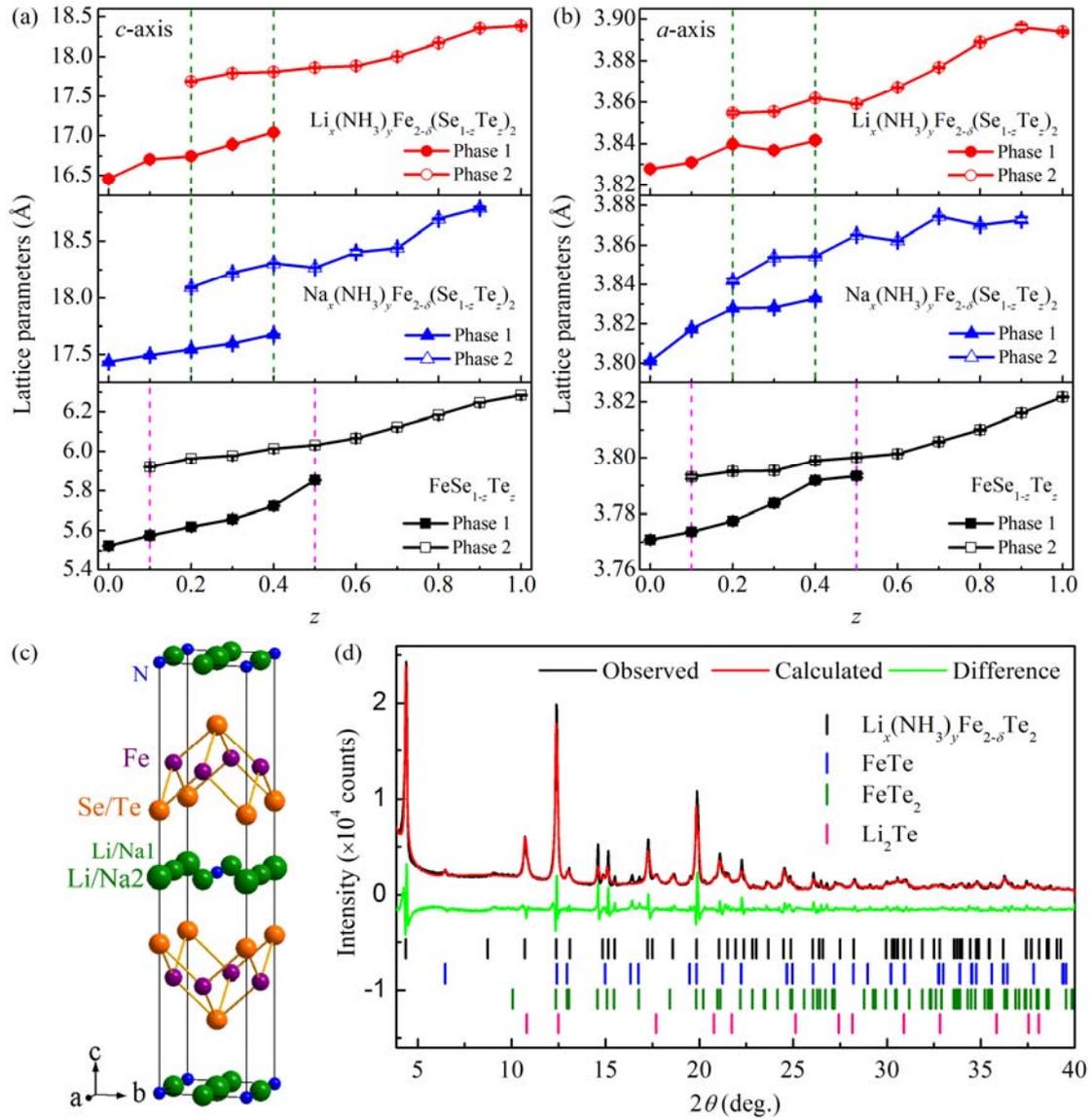

Fig.1 *Lei* et al.

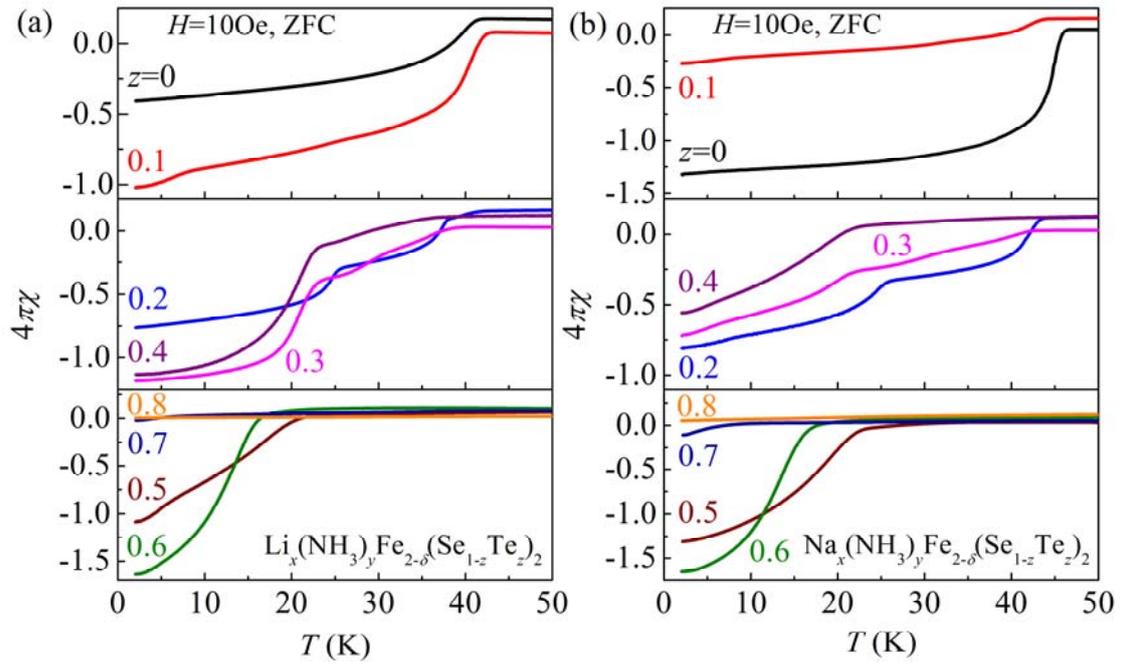

Fig. 2 *Lei* et al.

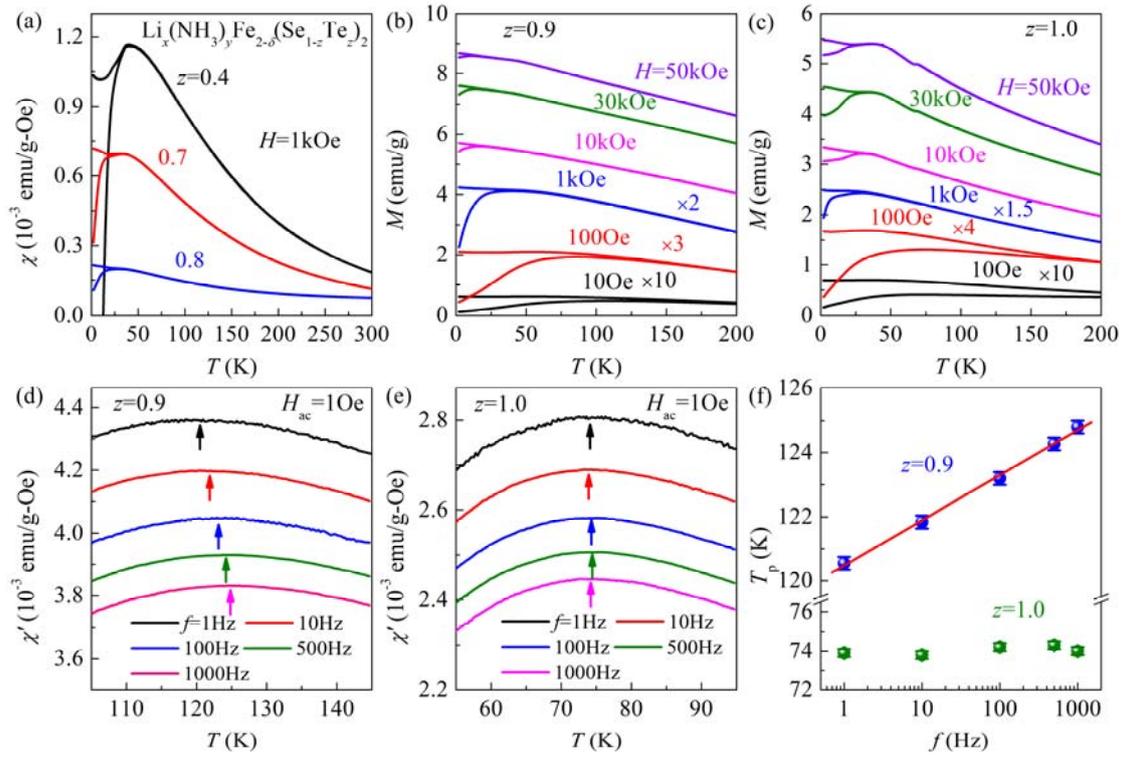

Fig. 3 *Lei* et al.

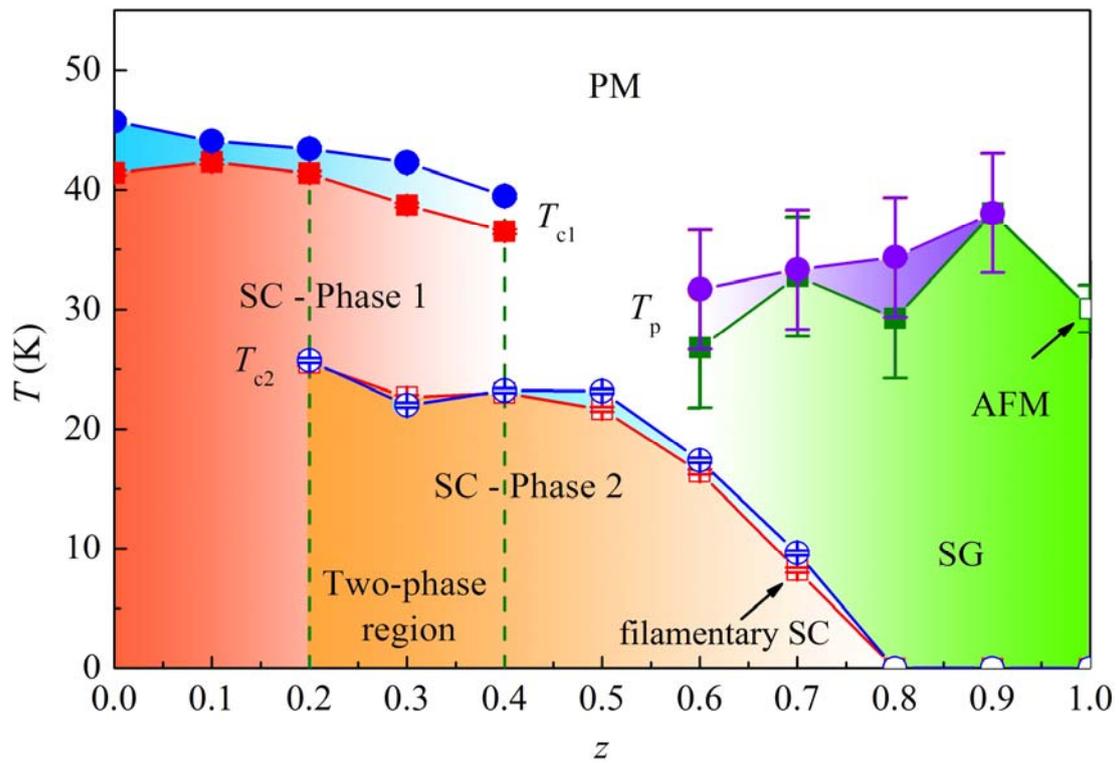

Fig. 4 *Lei* et al.

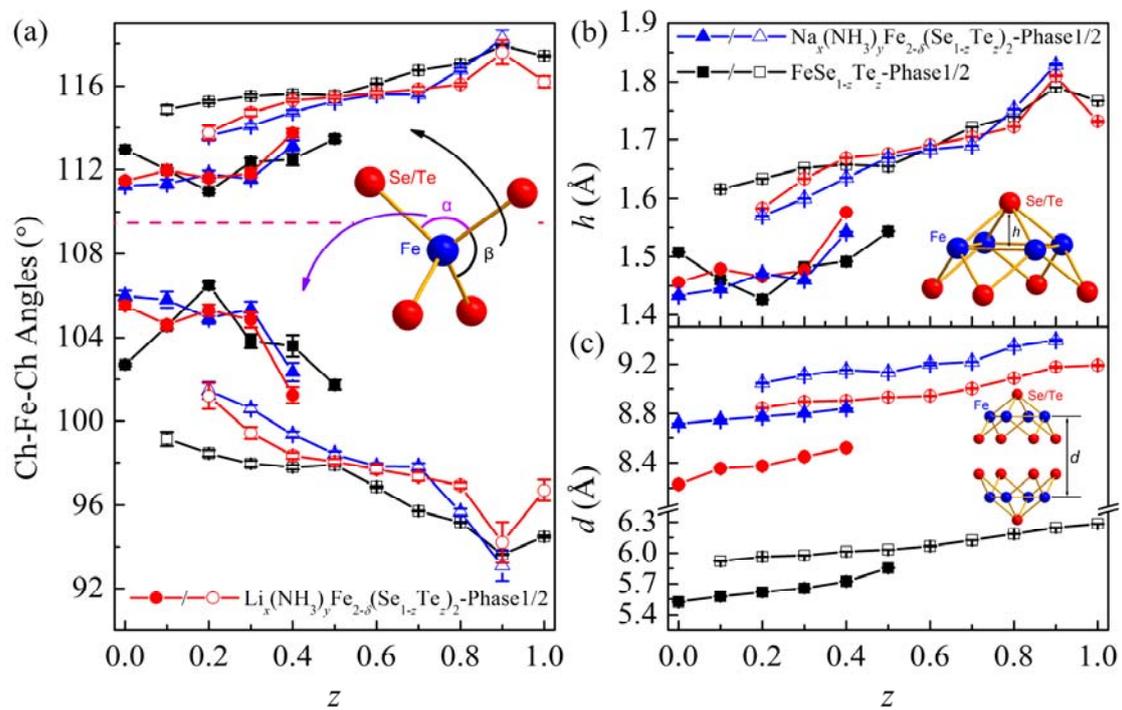

Fig. 5 *Lei* et al.

# Emergence of magnetism and controlling factors of superconductivity in Li/Na-ammonia co-intercalated FeSe$_{1-z}$Te$_z$

*Hechang Lei,*[1,†,§] *Jiangang Guo,*[1,†] *Fumitaka Hayashi,*[2] *and Hideo Hosono,*[1,2,3*]

[1] Materials Research Center for Element Strategy, Tokyo Institute of Technology, Yokohama 226-8503, Japan

[2] Frontier Research Center, Tokyo Institute of Technology, Yokohama 226-8503, Japan

[3] Materials and Structures Laboratory, Tokyo Institute of Technology, Yokohama 226-8503, Japan

**Supplemental Material**

[†]These authors contributed equally to this work.

§Present address: Department of Physics, Renmin University of China, Beijing 100872, China

*To whom correspondence may be addressed. E-mail: hosono@msl.titech.ac.jp

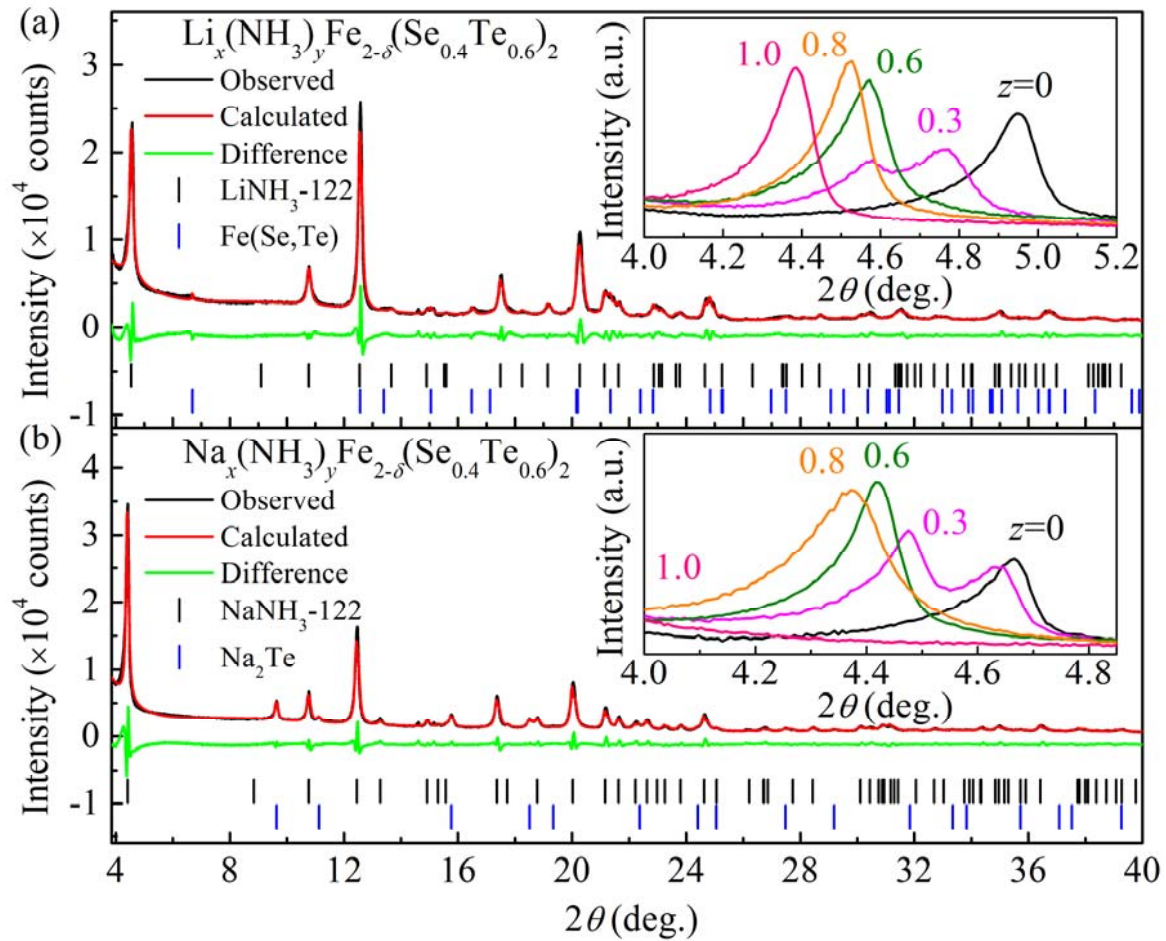

**Figure S1.** PXRD patterns and fitted results of (a) $Li_x(NH_3)_yFe_{2-\delta}(Se_{0.4}Te_{0.6})_2$ and (b) $Na_x(NH_3)_yFe_{2-\delta}(Se_{0.4}Te_{0.6})_2$. Other samples with various $z$ values exhibit similar PXRD patterns except for $Na_x(NH_3)_yFe_{2-\delta}Te_2$. The structure of $Li_{0.6}(ND_{2.8})Fe_2Se_2$ is used as an initial model and the structural parameters are obtained from Rietveld refinement. There are trace amounts of second phases like $(Li/Na)_2Te$, $Fe(Se, Te)_2$ and unreacted $Fe(Se, Te)$ existing in the intercalated samples. Insets: enlarged parts of PXRD patterns for (002) diffractions for $(Li/Na)_x(NH_3)_yFe_{2-\delta}(Se_{1-z}Te_z)_2$ with $z = 0, 0.3, 0.6, 0.8$, and $1.0$.

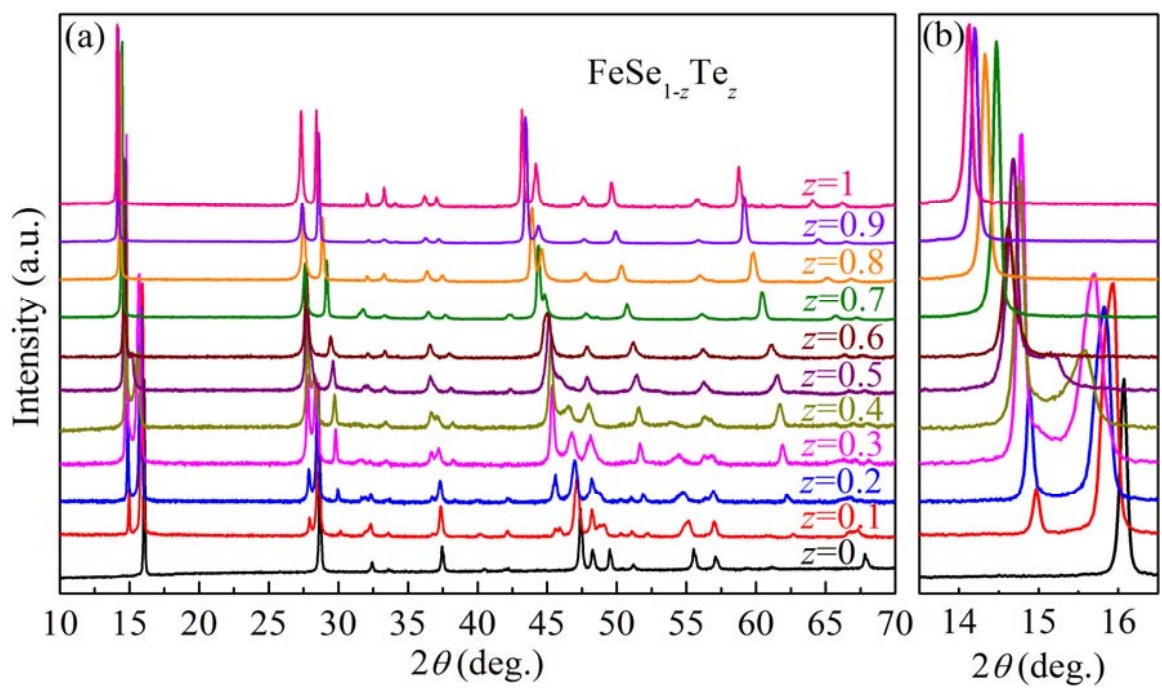

**Figure S2.** (a) PXRD patterns of FeSe$_{1-z}$Te$_z$. (b) Enlarged part of XRD patterns showing the (001) peak.

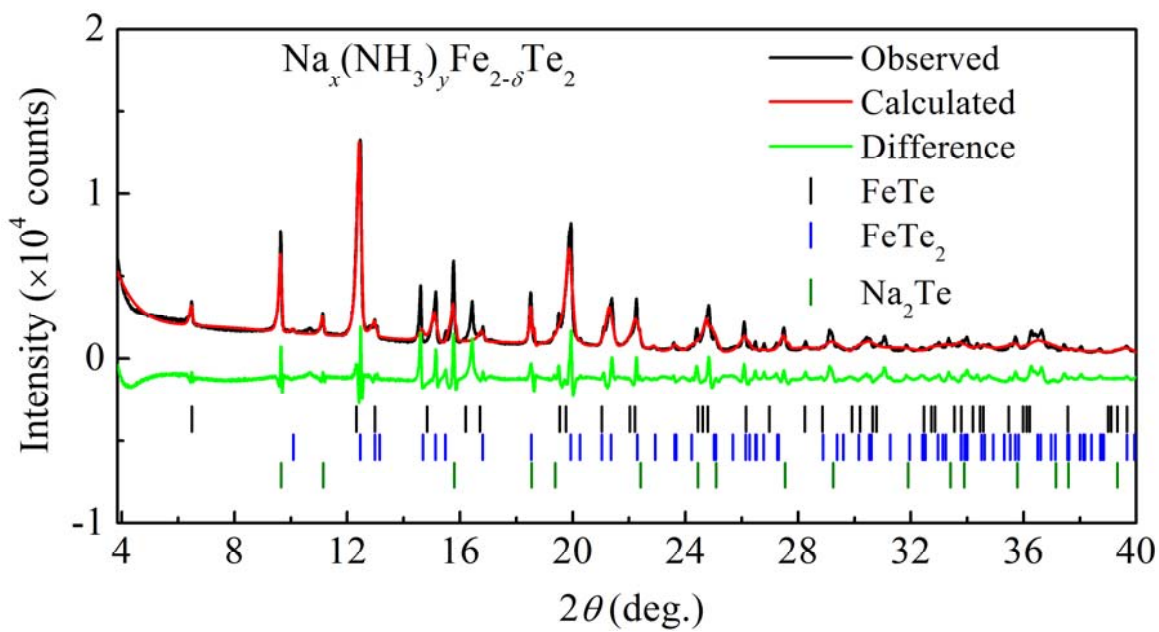

**Figure S3.** PXRD patterns of reaction products after immersing FeTe into Na-NH$_3$ solution. Only binary phases such as FeTe, FeTe$_2$ and Na$_2$Te appear.

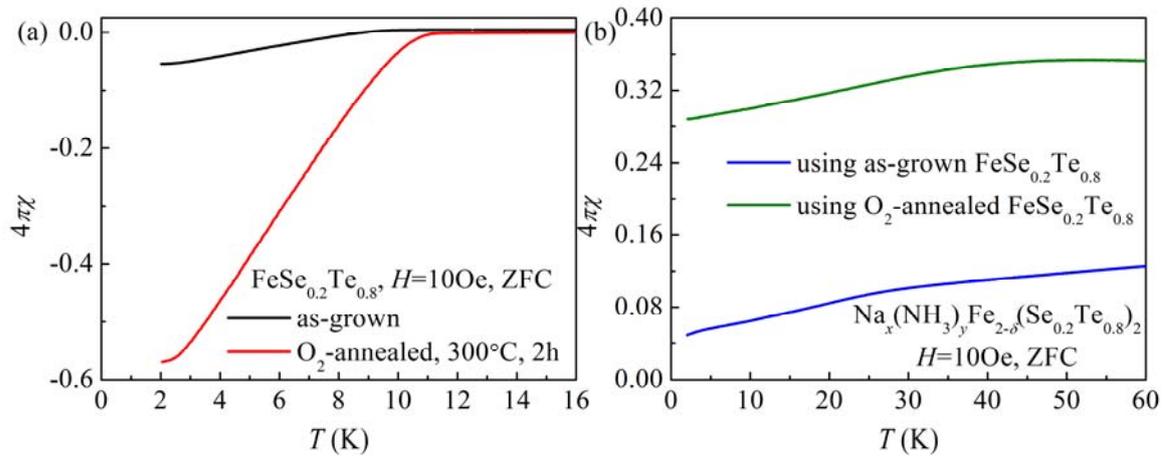

**Figure S4.** (a) Magnetic susceptibilities $4\pi\chi(T)$ of as-grown and $O_2$-annealed $FeSe_{0.2}Te_{0.8}$ and (b) $4\pi\chi(T)$ curves of $Na_x(NH_3)_yFe_{2-\delta}(Se_{0.2}Te_{0.8})_2$ prepared using as-grown and $O_2$-annealed $FeSe_{0.2}Te_{0.8}$ at $H$ = 10 Oe with zero-field-cooling (ZFC) mode.

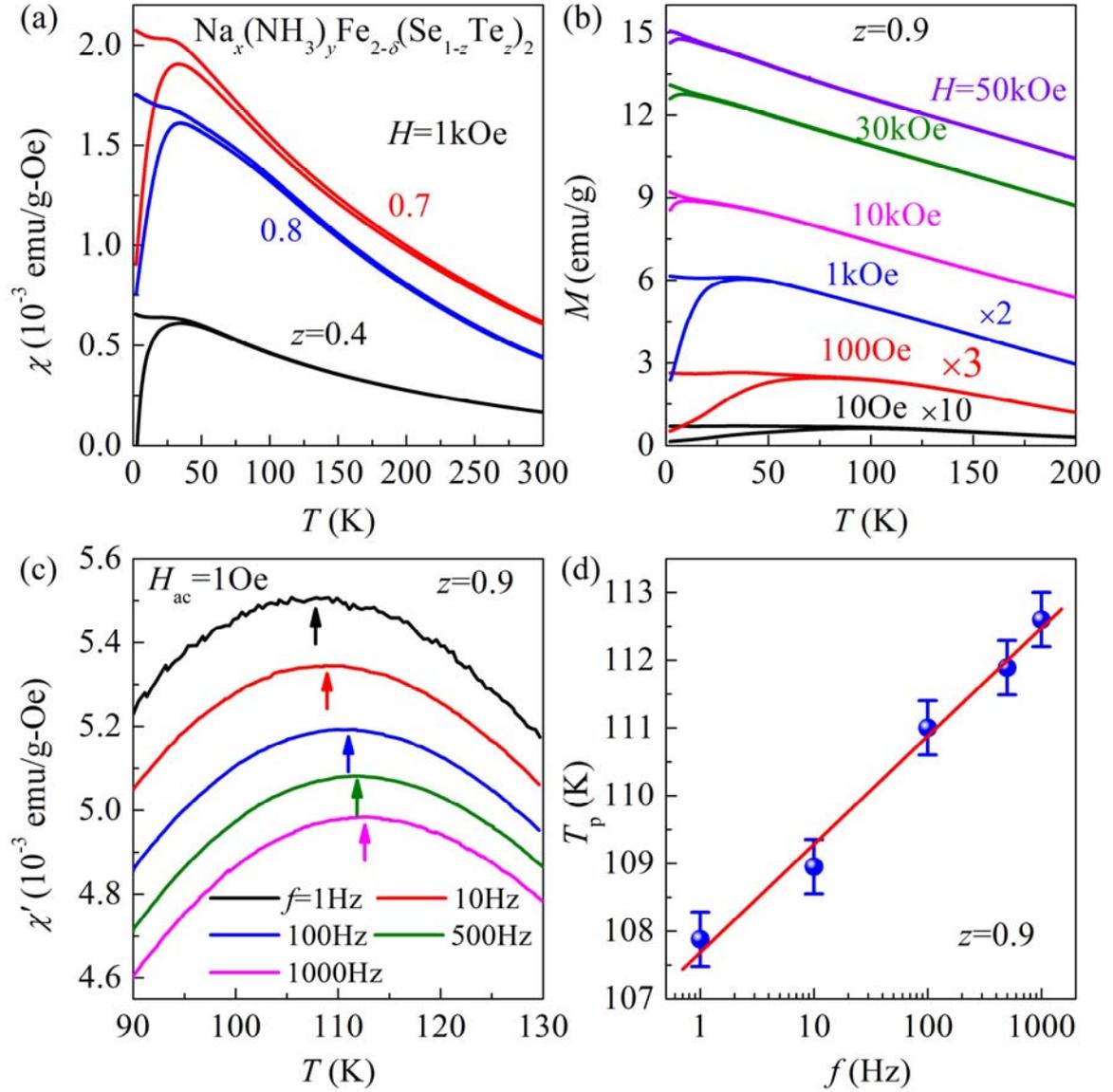

**Figure S5.** (a) Temperature dependence of the magnetic moment $M(T)$ at various fields. (b) The real part of ac susceptibility $\chi'(T)$ at $H_{ac}$ = 1 Oe with various frequencies, $f$. The arrows show the peak positions, $T_p$s, on the $\chi'(T)$ curves. (c) Frequency dependence of $T_p$. The solid line is the result of linear fit. The obtained $K$ (= $\Delta T_p /(T_p \Delta \log f)$) is 0.0117(3).

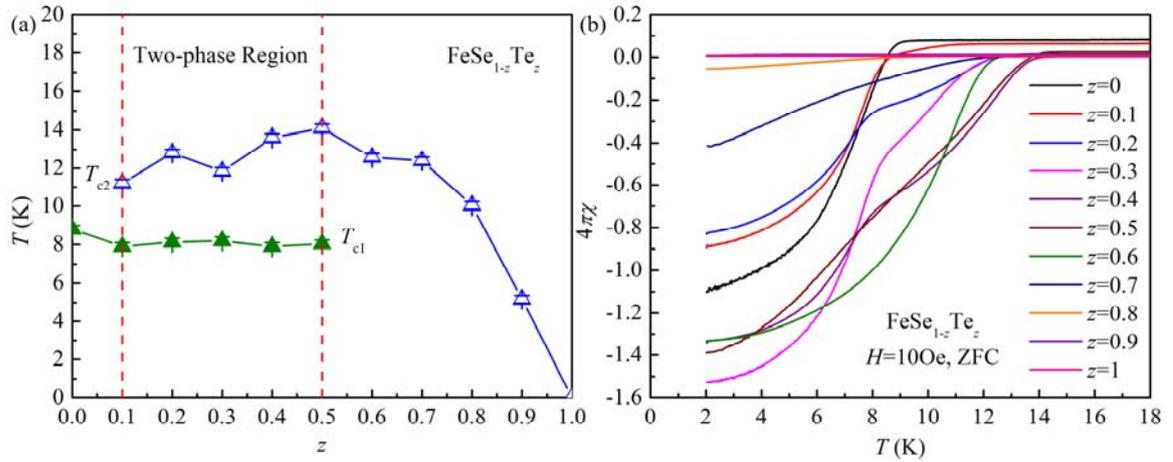

**Figure S6.** (a) Superconducting phase diagram of FeSe$_{1-z}$Te$_z$ determined from magnetic susceptibility measurements. The green solid and blue empty triangles represent two superconducting transition temperatures $T_{c1}$ and $T_{c2}$ of FeSe$_{1-z}$Te$_z$. (b) $4\pi\chi(T)$ curves of FeSe$_{1-z}$Te$_z$ measured at low-temperature range at $H$ = 10 Oe. For clarity, only the ZFC curves are shown.

**Table S1.** Nominal and actual average compositions of two series of $(Li/Na)_x(NH_3)_y Fe_{2-\delta}(Se_{1-z}Te_z)_2$ measured by energy-dispersive X-ray analysis (EDX) method.[*]

| Nominal z | $Li_x(NH_3)_y Fe_{2-\delta}(Se_{1-z}Te_z)_2$ | | $Na_x(NH_3)_y Fe_{2-\delta}(Se_{1-z}Te_z)_2$ | | |
|---|---|---|---|---|---|
| | Actual average $\delta$ | Actual average z | Actual average x | Actual average $\delta$ | Actual average z |
| 0.0 | 0.18(17) | 0.00 | 0.74(16) | 0.22(20) | 0.00 |
| 0.1 | 0.10(18) | 0.07(2) | 0.46(11) | 0.06(16) | 0.07(5) |
| 0.2 | 0.09(15) | 0.17(4) | 0.67(12) | 0.08(11) | 0.22(7) |
| 0.3 | 0.03(15) | 0.31(2) | 0.44(10) | 0.08(10) | 0.34(14) |
| 0.4 | 0.15(11) | 0.42(8) | 0.67(21) | 0.15(16) | 0.46(4) |
| 0.5 | 0.04(13) | 0.53(4) | 0.83(15) | 0.09(9) | 0.54(2) |
| 0.6 | 0.12(11) | 0.61(2) | 0.67(16) | 0.17(9) | 0.59(3) |
| 0.7 | 0.10(7) | 0.69(1) | 0.70(13) | 0.03(10) | 0.72(1) |
| 0.8 | 0.09(13) | 0.80(2) | 0.54(11) | 0.06(7) | 0.82(3) |
| 0.9 | 0.09(06) | 0.91(2) | 0.59(10) | 0.02(6) | 0.91(2) |
| 1.0 | 0.08(05) | 1.00 | - | - | - |

[*]The Li content in the intercalates cannot be determined by the EDX.

**Table S2.** Nominal and actual average z values in $FeSe_{1-z}Te_z$ measured by EDX method.

| Nominal z | Actual average z |
|---|---|
| 0.0 | 0.00 |
| 0.1 | 0.08(2) |
| 0.2 | 0.19(4) |
| 0.3 | 0.30(4) |
| 0.4 | 0.38(6) |
| 0.5 | 0.52(4) |
| 0.6 | 0.58(2) |
| 0.7 | 0.71(2) |
| 0.8 | 0.81(2) |
| 0.9 | 0.91(1) |
| 1.0 | 1.00 |